\documentclass[fleqn,usenatbib,useAMS]{mnras}

\usepackage{graphicx}	
\usepackage{amsmath}	
\usepackage{amssymb}	
\usepackage{multicol}        
\usepackage{bm}		
\usepackage{pdflscape}	

\usepackage{deluxetable}

\usepackage{hyperref}
\usepackage{lineno}

\usepackage{color}

\makeatletter
\def\@to{to}
\makeatother

\usepackage[T1]{fontenc}
\usepackage{ae,aecompl}

\usepackage{newtxtext,newtxmath}

\title[Causality distance feasibility]{Estimating the feasibility of `standard speed-gun' distances.}

\author[J. A. Hodgson]{Jeffrey A. Hodgson$^{1}$\thanks{Contact e-mail: jhodgson@sejong.ac.kr}, Benjamin L'Huillier$^{1}$, Ioannis Liodakis$^{2}$, Sang-Sung Lee$^{3,4}$,\newauthor Arman Shafieloo$^{3,4}$
\\
$^{1}$Department of Physics and Astronomy, Sejong University, 209 Neungdong-ro, Gwangjin-gu, Seoul, South Korea \\
$^{2}$Finnish Centre for Astronomy with ESO, University of Turku, Quantum, Vesilinnantie 5, FI-20014, Finland \\
$^{3}$Korea Astronomy and Space Science Institute, Daedeokdae-ro 776, Daejeon, South Korea \\
$^{4}$Korea University of Science and Technology, 217 Gajeong-ro, Yuseong-gu, Daejeon 34113, Korea 
}

\date{Last updated 2015 May 22; in original form 2013 September 5}

\pubyear{2022}

\begin{document}
\label{firstpage}
\pagerange{\pageref{firstpage}--\pageref{lastpage}}
\maketitle

\begin{abstract}
In a previous paper, we demonstrated a single-rung method for measuring cosmological distances in active galactic nuclei (AGN) that can be used from low redshift ($z < 0.1$) to high redshift ($z > 3$). This method relies on the assumption that the variability seen in AGN is constrained by the speed of light during a flare event and can therefore be used to estimate the size of an emitting region. A limitation of this method is that previously, the Doppler factor was required to be known. In this paper, we derive an extension of the `standard speed-gun' method for measuring cosmological distances that depends on the maximum intrinsic brightness temperature that a source can reach, rather than the Doppler factor. If the precise value of the intrinsic brightness temperature does not evolve with redshift and flares are statistically independent, we can in principle improve the errors on measurements of the matter content of the universe (in a flat $\Lambda$CDM model) statistically. We then explored how well a future observing program would constrain cosmological parameters. We found that recovering the input cosmology depends critically on the uncertainty of the intrinsic brightness temperature and the number of flares observed.
\end{abstract}







\section{Introduction}


\begin{figure*}
    \centering
    \includegraphics[width=\textwidth]{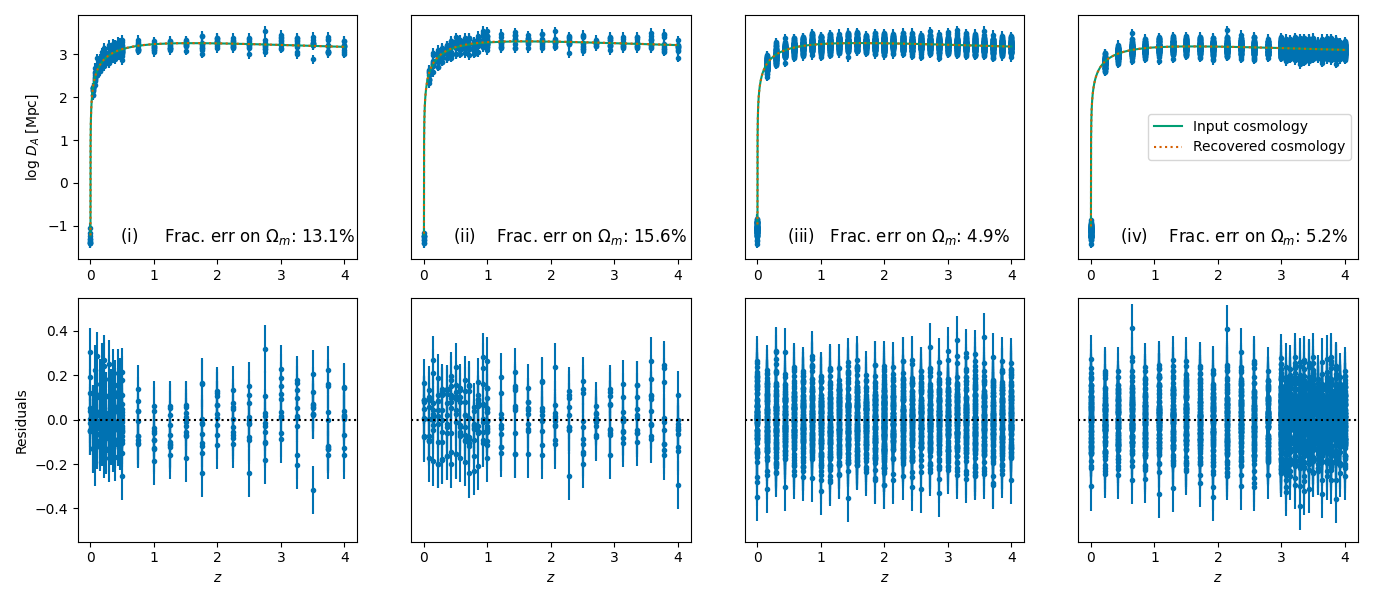}
    \caption{Top: Simulated observations, assuming 30 sources with 10 flares observed per source assuming a 25\% on the intrinsic brightness temperature. The fractional error on $\Omega_{\rm m}$ from the recovered cosmology is shown in the figures. Input cosmology in green. Recovered cosmology in orange. Simulated data in blue. Four redshift distributions for the simulated observations were made: i) half the sources between $0 < z < 0.5$ and the remaining half between $0.5 < z < 4$; ii) half the sources between $0 < z < 1$ and the remaining half between $1 < z < 4$; iii) as in case (ii) except pivoting on $z=2$ (i.e. equally distributed) and iv) as before except pivoting on $z=3$. Bottom: residuals (simulated data - input cosmology). }
    \label{fig:z_distr_compare}
\end{figure*}
In our previous paper \citep{hodgson2020} (hereafter Paper I), we demonstrated a new single-rung method for measuring cosmological distances that relies on the speed-of-light to calibrate a standard ruler in active galactic nuclei (AGN) that we call the 'standard speed-gun'. A common feature of AGN that have their jets pointed toward the observer are bright jets that exhibit special relativistic effects such as apparent superluminal motions, increased variability, and increased observed flux densities \citep[e.g.][]{mojave16,jor17,weaver22}. These effects are normally accounted for by estimating the relativistic Doppler factor, which is itself a function of the viewing angle to the source and its intrinsic Lorentz factor. In our previous paper, we made a well-justified assumption that the Doppler factor was of order unity in the source 3C 84 \citep[e.g.][]{liodakis_fgamma,jor17}. While this assumption was likely reasonable in 3C 84, for sources at higher redshifts, relativistic effects must be taken into account. Unfortunately, it would be expected that the Doppler factor will evolve with redshift, with more distant sources likely to be more Doppler boosted. This is because Doppler boosting makes sources appear brighter than they are intrinsically and therefore will be more likely to be detected at greater distances for a given instrumental sensitivity. 
A common way of estimating the Doppler factor is to assume that there is a maximum intrinsic brightness temperature $T_{\rm B, int}$ that the source can reach. There are several theoretical estimates have been made for this limit. The most commonly cited are the equipartition limit of $\sim 5 \times 10^{10}$\,K \citep{readhead1994}, $< 5 \times 10^{11}$\,K \citep{singal1986} and the inverse Compton limit $< 10^{12}$\,K \citep{kpt69}. Additionally, if the limit follows the equipartition limit, it can be calculated theoretically \citep{readhead2021}. Several observational studies have been performed in order to constrain the true value of $T_{\rm B, int}$. Estimates vary widely, including $\leq 10^{11}$\,K \citep{lahteenmaki99}, $\sim 2 \times 10^{10}$\,K \cite{cohen03}, $2.78 \pm 0.72 \times 10^{11}$\,K \citep{liodakis18} and $4.1 (\pm 0.6) \times 10^{10}$\,K \citep{mojaveTBint}. In the case of \citet{liodakis18}, only the brightest flares were used in the analysis and found that the $T_{\rm B, int}$ distribution was best described by a Gaussian: suggesting that there is not a bias towards lower or higher intrinsic brightness temperatures when using the brightest flares. In this paper, we derived a new expression for the angular diameter distance that depends on the intrinsic brightness temperature, rather than the Doppler factor. The exact value of $T_{\rm B, int}$ will affect measurements of the Hubble parameter (effectively the absolute scaling of the distance-redshift relationship), but so long as the value itself does not evolve with redshift and the observed flares are statistically independent, we can measure the matter content of the universe ($\Omega_{\rm m}$, in a flat $\Lambda$CDM cosmology) and improve the precision of the measurement statistically. 
\begin{figure*}
    \centering
    \includegraphics[width=\textwidth]{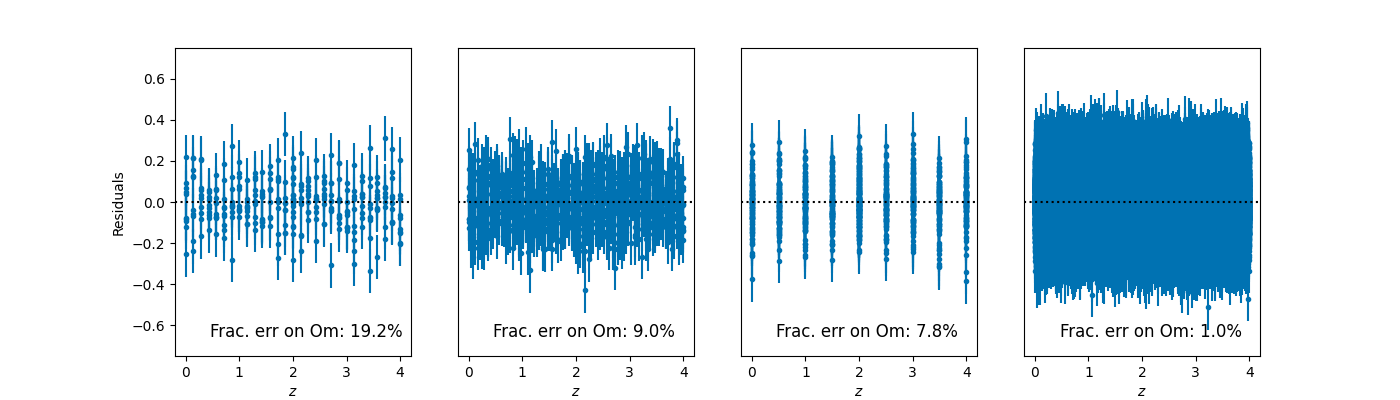}
    \includegraphics[width=\textwidth]{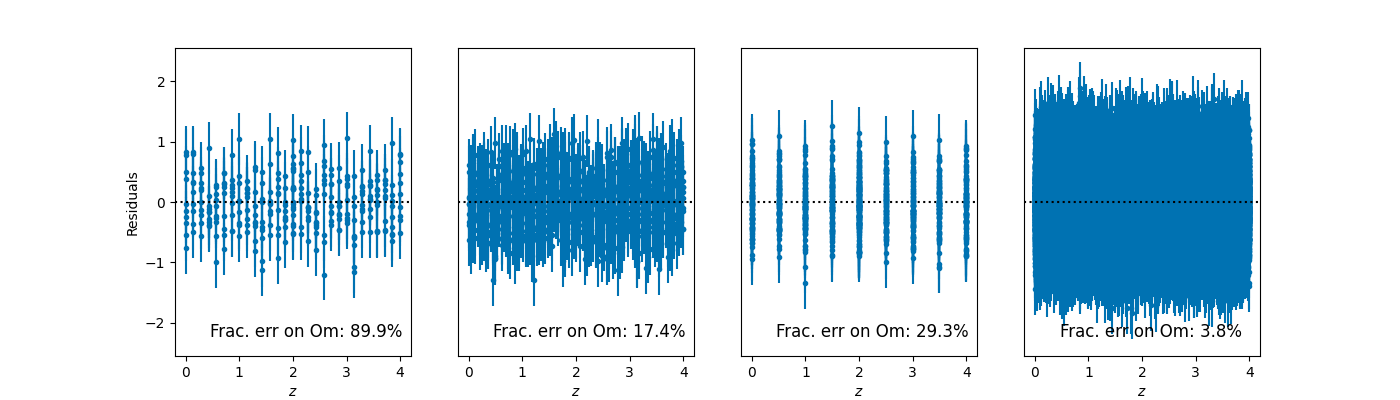}
    \caption{Residuals of simulated observations, exploring the effect of the number of sources, the number of flares observed per source and the uncertainty on $T_{\rm B,int}$. The fractional error on $\Omega_{\rm m}$ from the recovered cosmology is shown in the figures. Input cosmology in orange. Recovered cosmology in green. Simulated data in blue. Four scenarios are: i) 30x sources, 10x flares (as above); ii) 100x sources, 10x flares; iii) 10x sources, 100x flares and iv) 1000x sources, 100x flares. Top panel assumes 25\% uncertainty on $T_{\rm B,int}$, bottom panel assumes 100\% uncertainty on $T_{\rm B,int}$.}
    \label{fig:num_sources_tbint_compare}
\end{figure*}
We begin by defining the brightness temperature under the Rayleigh-Jeans approximation \citep[e.g.][]{kovalev05,kang21}:
\begin{equation}\label{tb_initial}
    T_{\rm B} = \frac{I_{\nu} c^{2}}{2 k_{\rm B} \nu^{2}},
\end{equation}
where $I_{\nu}$ is the intensity at a given frequency $\nu$ and $k_{\rm B}$ is Boltzmann's Constant. In the case of VLBI observations, the intensity can be observed as $S/\Omega$, where $\Omega$ is the solid angle of the emitting region on the sky and $S$ is the observed flux density. Typically in VLBI, we fit a circular Gaussian to the image. Therefore, the solid angle can be expressed as $\Omega = \pi \theta^{2} /4\ln{2}$, where $\theta_{\rm VLBI}$ is the FWHM of the circular Gaussian. \\
The observed quantities in the source frame (primed) are affected by cosmological redshift and in the following ways: $\nu = \nu' / (1+z) $; $S = S' / (1+z)$ and $\theta_{\rm VLBI} = \theta_{\rm VLBI}'$. Relativistic corrections may also be required. They are transformed in the following ways (denoted by $^{*}$) $\nu = \nu^{*}\delta $; $S = S^{*}\delta^{3}$ and $\theta_{\rm VLBI} = \theta_{\rm VLBI}^{*}$. This leads to the observed source-frame brightness temperature from VLBI observations (assuming a single resolved component with a flat spectrum):
\begin{equation}\label{TB_VLBI}
    T_{\rm B,VLBI}'^{*} = \frac{2 \ln{2} c^{2} \delta S (1+z)}{\pi k_{\rm B} \nu^{2} \theta_{\rm VLBI}^{2}}.
\end{equation}
This means that the observed VLBI brightness temperature is related to the true intrinsic brightness temperature, $T_{\rm B,int}$, as $T_{\rm B,VLBI}' = \delta T_{\rm B,int}$ or $T_{\rm B,VLBI} = T_{\rm B,int} \delta / (1+z) $.
An alternative way to measure the brightness temperature is to assume that the size of an emitting region is constrained by the speed of light. This allows us to estimate a ``variability size'' (as derived in \citet{hodgson2020}), and including $\Delta t$ relativistically transformed as $\Delta t = \Delta t^{*}/\delta$:
\begin{equation}\label{theta_var}
    \theta_{\rm var}'^{*} = \frac{c \delta \Delta t}{(1+z) D_{\rm A}},
\end{equation}
where $\Delta t$ is the variability timescale and $D_{\rm A}$ is the angular diameter distance to the source. 
We can then substitute Eq. \ref{theta_var} into Eq. \ref{TB_VLBI} (including the relativistic corrections) to derive the observed source-frame variability brightness temperature \citep[e.g.][]{liodakis_fgamma}:
\begin{equation}
    T_{\rm B,var}'^{*} = \frac{2 \ln{2} D_{\rm A}^{2} \delta^{3} (1+z)^{3} \Delta S }{\pi k_{\rm B} \nu^{2} \Delta t^{2}}.
\end{equation}
where $\Delta S$ is the peak flux density of a flare. This should have the same value as $S$ if the variability information is derived from the same VLBI data. Furthermore, because $\Delta S$ is a measure of the difference in flux density measured from the beginning to the peak of a flare, it follows that if $\Delta S = S$, then the VLBI flux density measurement should be taken at the peak of a flare. This also means that the observed \emph{variability} brightness temperature is related to the true intrinsic brightness temperature, $T_{\rm B,int}$, as $T_{\rm B,var}' = \delta^{3} T_{\rm B,int}$ or $T_{\rm B,var} = T_{\rm B,int} \delta^{3}/(1+z)^{3} $.
Due to the differing dependencies on $\delta$, we can therefore combine these observations to derive direct estimates of both the relativistic Doppler factor (in either the source or observer frames):
\begin{equation}\label{Dopp_eq}
    \delta = \sqrt{\frac{T_{\rm B,var}'}{T_{\rm B,VLBI}'}}
\end{equation}
and the intrinsic brightness temperature:
\begin{equation}
    T_{\rm B,int} = \frac{T_{\rm B,VLBI}'^{3/2}}{T_{\rm B,var}'^{1/2}}.
\end{equation}
Since there is a dependence on the distance, these expressions can be rearranged to measure the distance in terms of the Doppler factor:
\begin{equation}\label{DA_Delta}
    D_{\rm A} = \frac{c \Delta t \delta}{\theta_{\rm VLBI} (1+z)}
\end{equation}
or equivalently in terms of the intrinsic brightness temperature:
\begin{equation}\label{DA_TBint}
    D_{\rm A} = \frac{2 \ln{2} c^{3} S \Delta t}{\pi k_{\rm B} T_{\rm B,int} \nu^{2} \theta_{\rm VLBI}^{3}}.
\end{equation}
A scaling factor of either 1.6x or 1.8x should be included in Eq. \ref{DA_Delta} in order to convert the measured Gaussian from the VLBI observations to either a disk-like or sphere-like geometry respectively. As in \citet{hodgson2020}, we adopted a compromise 1.7x scaling, with the uncertainty included in the error budget. In the case of Eq. \ref{DA_TBint}, no scaling is included as the equation still includes the Gaussian assumption. 
\section{Discussion}
There are some interesting features of Eq. \ref{DA_TBint}. The first is that it depends on the intrinsic brightness temperature $T_{\rm B,int}$ rather than the Doppler factor. So long as $T_{\rm B,int}$ does not evolve with redshift, we can improve measurements of the curve of the distance-redshift statistically. The exact value of $T_{\rm B,int}$ is expected to either not vary with redshift, or weakly if the limit follows the equipartition limit. We note that if $T_{\rm B,int}$ follows the theoretical equipartition limit, we can calculate the expected value \citep{readhead1994,readhead2021}. Another potential problem could be flare-to-flare variations in $T_{\rm B,int}$, leading to biases in the $T_{\rm B,int}$ estimates. However, so long as the brightest flares are used, this may not be an issue \citep{liodakis18}. Additionally, a bias like this would not be expected to evolve with redshift, therefore affecting $H_{0}$ measurements and not $\Omega_{\rm m}$. Another interesting feature is that there is no redshift dependence. This means that we could roughly estimate redshifts for sources without redshift measurements if a cosmology is assumed. 
In the published observational parameters for 3C\,84 from \citet{hodgson2020}, we calculated the Doppler factor and $T_{\rm B,int}$ assuming both of the most recent estimates of the Hubble Constant from SH0ES \citep{shoesH0} and from Planck \citep{planckH0}. These values are shown in Table \ref{Dops_TBint}. The Doppler factor appears to be consistent with unity, within errors in both cases. The $T_{\rm B,int}$ estimates are consistent with the value determined by \citet{liodakis18}. It is important to note that while it would be expected that the Doppler factor evolves with redshift, we believe it is unlikely that $T_{\rm B,int}$ would, or at least in a different way. 



In order to explore how this would potentially affect our observations, we made some simple simulations of a future observing program. The key assumptions are that the flares are statistically independent and that $T_{\rm B,int}$ does not evolve with redshift. Blazar flaring is known to be a stochastic process and well modeled by damped random walk processes \citep[e.g.][]{kowlowski2016}. While this doesn't necessarily mean that the flares are statistically independent, it does hint in that direction. In Fig. \ref{fig:z_distr_compare}, we first explored how the redshift distribution affects the fractional uncertainty on $\Omega_{\rm m}$. We assumed an initial sample of 30 sources over a redshift range of $0 < z < 4$. We then distributed over the redshift range in four ways. They were i) half the sources between $0 < z < 0.5$ and the remaining half between $0.5 < z < 4$; ii) half the sources between $0 < z < 1$ and the remaining half between $1 < z < 4$; iii) as in case (ii) except pivoting on $z=2$ (i.e. equally distributed) and iv) as before except pivoting on $z=3$. Assuming a fractional uncertainty of 25\% on $T_{\rm B,int}$ \citep{liodakis18}, we found that having more sources at higher redshift tends to improve the fractional error $\Omega_{\rm m}$, although the uniformly distributed sources scenario recovered $\Omega_{\rm m}$ roughly as well as when the sources were concentrated at high redshift. For context, the fractional error on $\Omega_{\rm}$ from the Planck Collaboration is $\sim$2.2\% \citep{planckH0}, and the error from the latest supernovae catalogs is $\sim$5\% \citep{pantheon2022}. Additionally, we note that in this case, we would be more interested in observing deviations from the expected cosmology at high-z \citep[e.g.][]{zhao_dynamical_darkenergy_2018,risaliti19}. 
We then explored how the number of observations and the uncertainty on $T_{\rm B,int}$ affects our recovered cosmology. We tested four observational scenarios (all with sources equally distributed over the redshift range): i) 30x sources, 10x flares (as above); ii) 100x sources, 10x flares; iii) 10x sources, 100x flares and iv) 1000x sources, 100x flares. Observing 100x flares per source is likely not realistic, but allows us to explore if it is better to have more sources or simply more observations. We included the last case as an example of a very large observing program. We then compared these scenarios for two uncertainties on $T_{\rm B,int}$ of 25\% (as above) and 100\% (a "worst-case scenario"). The results are shown in Fig. \ref{fig:num_sources_tbint_compare}. It can be seen that it is the total number of flares observed that is important. It does not appear to depend if it is a smaller number of sources with more flare per source or simply more sources. Given that there may only be $\sim$1 flare per source, per year, this suggests that observing more sources would be a more efficient way of improving the error budget. We can also see that the uncertainty on $T_{\rm B,int}$ critically affects the recovered cosmology. We can see that if $T_{\rm B,int}$ is known to 25\%, we are competitive with other methods \citep[e.g.][]{pantheon2022}, with  realistic observing programs, such as in scenarios (ii) and (iii). If $T_{\rm B,int}$ is not known well, we would need a much larger observing program in order to be competitive. \citet{angelakis_fgamma} suggests that 1-2 flares per year per source is reasonable, and therefore the method could be competitive within 5-10 years of observations. 
In principle, we can also fit $T_{\rm B,int}$ as a free parameter with the other cosmological parameters. But $D_{\rm A} \propto 1/ T_{\rm B,int}$ and also $D_{\rm A} \propto 1/H_{0}$, which means we can only constrain the ratio. However, with a cosmology-independent measure of the Doppler factor, the degeneracy could be broken. A good candidate for a cosmology-independent measure of the Doppler factor is the inverse-Compton Doppler factor \citep{ghisellini93,liodakis17}. Note that this does not have to be estimated for every source, but a smaller subset would be sufficient. That the Doppler factor evolves with redshift is not necessarily a problem, so long as the \emph{correction} does not evolve with redshift. Furthermore, an observational program such as this would provide an excellent sample to investigate AGN evolution.
\section{Conclusions}
In this paper, we have derived a new expression for the angular diameter distance for blazars that depends on a maximum intrinsic brightness temperature ($T_{\rm B,int}$), rather than the Doppler factor. In the case of 3C\,84, we assumed two cosmologies and derived estimates of the Doppler factor and $T_{\rm B,int}$. The Doppler factor was found to be consistent with unity and $T_{\rm B,int}$ was consistent with the independent estimate of \citet{liodakis18}. We explored how well a future observing program could constrain cosmological parameters in a flat $\Lambda$CDM model. We found that recovering the input cosmology depends critically on the uncertainty on $T_{\rm B,int}$ and the number of flares observed.
In our next paper, we will investigate if $T_{\rm B,int}$ evolves with redshift.
\begin{table}
	\centering
	\caption{Estimates of the Doppler factor and $T_{\rm B,int}$ in 3C 84, with different Hubble Constant estimates.}
	\label{Dops_TBint}
	\begin{tabular}{ccc} 
		\hline
		Hubble Constant & Doppler factor & $T_{\rm B,int}$  \\ 
		 km s Mpc & -  &  [$\times 10^{11}$K]  \\
		\hline
		67.40 $\pm$ 0.50 & $1.07_{0.13}^{0.13}$ & $4.06^{1.07}_{1.62}$  \\
		73.04 $\pm$ 1.04 & $0.99^{0.12}_{0.12}$ & $4.44^{1.17}_{1.70}$  \\ 
		\hline
	\end{tabular}
\end{table}




\bibliographystyle{mnras}

\textbf{Acknowledgements:} JH would like to acknowledge the support by National Research Foundation of Korea 2021R1C1C1009973. AS would like to acknowledge the support by National Research Foundation of Korea NRF2021M3F7A1082053, and the support of the Korea Institute for Advanced Study (KIAS) grant funded by the government of Korea. BL would like to acknowledge the support of the National Research Foundation of Korea (NRF-2019R1I1A1A01063740) and (NRF-2022R1F1A1076338). S-SL was supported by the National Research Foundation of Korea (NRF) grant funded by the Korea government (MIST) (2020R1A2C2009003).

\textbf{Data availability:} The data underlying this article will be shared on reasonable request to the corresponding author.




\bsp	
\label{lastpage}
\end{document}